# Surface and Bulk Structural Properties of Single Crystalline $Sr_3Ru_2O_7$

Biao Hu[1], Gregory T. McCandless[1], Melissa Menard[2], V. B. Nascimento[1], Julia Y. Chan[2], E. W. Plummer[1], R. Jin[1,*]

[1]Department of Physics and Astronomy, Louisiana State University, Baton Rouge, LA 70803
[2]Department of Chemistry, Louisiana State University, Baton Rouge, LA 70803

## Abstract

We report temperature and thermal-cycling dependence of surface and bulk structures of double-layered perovskite $Sr_3Ru_2O_7$ single crystals. The surface and bulk structures were investigated using low-energy electron diffraction (LEED) and single-crystal X-ray diffraction (XRD) techniques, respectively. Single-crystal XRD data is in good agreement with previous reports for the bulk structure with $RuO_6$ octahedral rotation, which increases with decreasing temperature (~ 6.7(6)° at 300 K and ~ 8.1(2)° at 90 K). LEED results reveal that the octahedra at the surface are much more distorted with a higher rotation angle (~ 12° between 300 and 80 K) and a slight tilt ((4.5±2.5)° at 300 K and (2.5±1.7)° at 80 K). While XRD data confirms temperature dependence of the unit cell height/width ratio (*i.e.* lattice parameter *c* divided by the average of parameters *a* and *b*) found in a prior neutron powder diffraction investigation, both bulk and surface structures display little change with thermal cycles between 300 and 80 K.

The ruthenate Ruddlesden-Popper (RP) series $Sr_{n+1}Ru_nO_{3n+1}$ (n = 1, 2, 3, …∞) exhibit rich electronic and magnetic properties covering a range from diamagnetic superconductor[1] (n = 1) to paramagnetic conductor with antiferromagnetic (AFM) correlation[2] (n = 2) to ferromagnetic (FM) metal[3,4] (n = 3, ∞). Extensive studies on the single-layered (n = 1) $Sr_2RuO_4$ and the isovalently doped $(Sr,Ca)_2RuO_4$ system reveal that the lattice degree of freedom plays a critical role in their physical properties, both in bulk[5,6] and on the surface[7,8,9]. Theoretical calculations[10] also indicate that the rotation and tilt of $RuO_6$ octahedra in $(Sr,Ca)_2RuO_4$ are closely coupled to the ferromagnetism and antiferromagnetism, respectively. Therefore, precise determination of the structural properties of the RP series is the key towards understanding of their exotic physical properties.

The motivation for investigating the structural properties of double-layered (n = 2) $Sr_3Ru_2O_7$ is multifold. First of all, it displays unique physical properties, different from the rest of the RP series. Although there is no long-range magnetic ordering under ambient pressure, a short-range AFM-type correlation develops below ~ 20 K, as probed by magnetic susceptibility[2,11] and neutron scattering measurements[12]. Application of hydrostatic and uni-axial pressure can drive the system into a ferromagnetically ordered state[11,13]. The application of magnetic field leads to similar result – the system undergoes a metamagnetic transition[14]. These clearly indicate that the system has two competing magnetic interactions (AFM versus FM), which are coupled with the structural properties. Second, the bulk crystal structure of $Sr_3Ru_2O_7$ has been modeled after data collection with various diffraction techniques into three different space groups since the first report in 1990[15]. The space groups reported include *I*4/*mmm* (tetragonal, No. 139)[15-17], *Pban* (orthorhombic, No. 50)[2], and *Bbcb* (orthorhombic, No. 68)[18-21]. Third, according to previous neutron powder diffraction work[19], the refined structural parameters depend not only on temperature but also thermal cycling. The latter is yet to be verified. Fourth, the surface structural properties of $Sr_3Ru_2O_7$ have so far not been reported, although there was some information provided by scanning tunneling microscopy (STM)[22]. Creating a surface by cleaving a single crystal creates an environment with broken translational symmetry, which can lead to different structure at the surface[9,11]. In this article, we present the results of our low-energy electron diffraction (LEED) and single-crystal X-ray diffraction (XRD) measurements on $Sr_3Ru_2O_7$ single crystals.

Single crystals of $Sr_3Ru_2O_7$ were grown using the floating-zone technique, which is proven for producing high-quality crystals. A detailed description of the growth procedure of $Sr_3Ru_2O_7$ single crystals can be found in our previous publication[12]. The single crystals used in our experiments are of excellent quality as mosaicity ranges from 0.427(6)° to 0.482(7)°. For single-crystal XRD measurements, a single crystal of $Sr_3Ru_2O_7$, with approximate size of 0.03 x 0.08 x 0.08 $mm^3$, was selected and mounted with epoxy on a thin glass fiber attached to a brass fitting. After allowing sufficient time for the epoxy to dry and harden, vacuum grease was carefully applied at the adhesive intersection of the single crystal and the glass fiber. The combination of epoxy and vacuum grease was needed to provide the stability of the sample through the thermal cycles. The XRD experiment was conducted on a Nonius KappaCCD X-ray diffractometer with a Mo Kα radiation source ($\lambda$ = 0.71073 Å), a graphite monochromator, and an Oxford Cryosystems 700 series cryostream controller. Data collections were made at three different temperatures (298 K, 200 K, and 90 K with a cooling/warming rate of 5 K/minute) after waiting about 30 minutes for the temperature of the single crystal to stabilize and about 30 minutes for preliminary unit cell/crystal quality determination, diffraction limit estimation, and setup of the appropriate scan-set strategy using Nonius SuperGUI software. At each fixed temperature, data collections were approximately 1½ hours long, covering angle theta range of 1.0° to 27.5°. A lower monoclinic symmetry, 2/*m*, was used in order to increase the number of images collected for refinement. Each successive thermal cycle was completed in the following order: 1) data collection at 298 K, 2) lowering down temperature to 200 K, 3) data collection at 200 K, 4) lowering down temperature to 90 K, 5) data collection at 90 K, 6) warming up temperature to 200 K, 7) data collection at 200 K, and 8) warming up temperature to 298 K. This order of events was repeated without any delay between cycles. After the data collections were completed, the data refinement was done using the maXus package with SHELXL-97[23] and SIR97[24] software. Final refinement was completed using WinGX[25] with SHELXL-97. Missing symmetry was checked using the "ADDSYM" test in the PLATON[26] program.

For the LEED experiment, a $Sr_3Ru_2O_7$ single crystal was cleaved *in situ* under an ultrahigh-vacuum (UHV) chamber with a base pressure of $2\times10^{-9}$ Torr, producing a shiny and flat (001) surface. After cleaving at room temperature, the sample was immediately transferred into a $\mu$-metal shielded LEED chamber with a base pressure of $7.0\times10^{-11}$ Torr. The sample position was adjusted to reach a normal incidence condition for the primary electron beam. For thermal cycle experiments, LEED data was first collected at 300 K, 200K, 80K during cooling and then 200 K, 300 K during warming. At each setting temperature, the LEED pattern was collected within an energy range of 60 – 600 eV using a home-built video-LEED system. *I-V* curves, which are based on the intensity of the diffraction spots as a function of the energy of the primary electron beam, were generated from digitized diffraction patterns and subsequently normalized to the incident electron beam current, then numerically smoothed with a weighted five-point-averaging method. For convenience, the indices of the LEED pattern at the surface is based on orthorhombic space group, considering the rotation of octahedral $RuO_6$ in bulk (see below). All *I*-*V* curves were obtained by averaging symmetrically equivalent beams. Ten *I*-*V* curves [(1,1), (2,0), (2,2), (3,3), (4,0), (4,4), (1,2), (1,3), (1,4) and (3,0)] were collected between 300 K and 80 K.

Table I shows the bulk crystallographic parameters for T = 298, 200, and 90 K from the first thermal cycle. The $Sr_3Ru_2O_7$ crystal structure is best modeled with the tetragonal space group *I*4/*mmm* (No. 139) with Sr1 (4/*mmm*), Sr2 (4*mm*), Ru (4*mm*), O1 (4/*mmm*), O2 (4*mm*), and O3 (*m*). Fig. 1(a) is the bulk unit cell representation of $Sr_3Ru_2O_7$. The structure consists of two layers of corner-sharing $RuO_6$ octahedra interleaved with SrO layers, *i.e.*, $SrO(SrRuO_3)_n$ ($n = 2$). The Ru atoms are located in the center of each octahedron with the crystallographic *c*-axis is shown in the vertical direction. Because of the large atomic displacement parameters found for the equatorial oxygen atoms, we allowed the atomic position and the occupancy of O3 to be refined. The results of this refinement lead to changing the Wyckoff position from 8*g* to 16*n* and the split occupancy for O3, as used in previous *I*4/*mmm* models of $Sr_3Ru_2O_7$ by others[2,18]. The atomic displacement parameters for data collected at 298 K yielded a 75% reduction in the anisotropic parameter, $U^{22}$, of O3 after decreasing the occupancy of the 8*g* site to 0.5 and refining the previously fixed atomic *y* coordinate, as shown in Table II. At room temperature, the new O3 position (16*n* site) is 0.23 Å away from the mirror plane corresponding to an octahedral rotation of 6.7(6)° (see Fig. 1(b)), which is in good agreement with the rotational angles reported in Refs. 2, 18, and 20. However, both the reported structural models using neutron powder diffraction adopt a lower symmetry (orthorhombic) space groups to model $Sr_3Ru_2O_7$ (*i.e. Pban*[2] and *Bbcb*[18,20]). Looking for the superlattice or weak reflections that might justify lowering the symmetry to one of these reported orthorhombic space groups, two other single crystals were examined with longer data collection. The absence of superlattice intensities in our XRD data and the ability to model the octahedral rotation with split occupancy of the equatorial O3 atoms allow us to describe the bulk $Sr_3Ru_2O_7$ structure with the higher symmetry space group, *I*4/*mmm*, instead of *Pban* or *Bbcb*. Attempts to model our XRD data with an orthorhombic space group result in divergence of the refinement and/or warnings of missing symmetry when evaluated with PLATON[26]. Application of space group transformations were needed to resolve the missing symmetry and ultimately led to modeling the data with the tetragonal space group, *I4/mmm*. The data collected at three different temperatures, as shown in Tables I-III, converge with $R_1 \sim 3\%$ and a final difference map of < 2 eÅ$^{-3}$ with well-behaved atomic displacement parameters.

Table IV provides selected interatomic distances of $Sr_3Ru_2O_7$. As the temperature is decreased from 298 to 90 K, Ru-O1 (inner apical oxygen) bond distance increases from 2.0195(11) to 2.0263(10) Å. The Ru-O2 (outer apical oxygen) distance and the Ru-O3 (equatorial oxygen) distance do not statistically change within this temperature range. The Jahn-Teller distorted $Ru^{4+}$ ($d^4$) observed in the first single crystal structure report[14] is also present with the bond distance from Ru to the equatorial oxygens less than the bond distances from Ru to both the inner and outer apical oxygens. The octahedra are also slightly distorted as evidenced by the small symmetrical buckling of the bond angles for O1-Ru-O3 (slightly less than 90˚) and O2-Ru-O3 (slightly greater than 90˚). The difference in O1-Ru-O3 and O2-Ru-O3 bond angles becomes smaller while lowering temperature. This indicates that the structure is less buckled at lower temperatures. Fig. 1(b) shows a top view of a $RuO_6$ octahedron at 298 K illustrating the O3-Ru-O3 bond angles with a rotational angle of 6.7(6)˚ off the mirror plane of the 2*mm* position. This $RuO_6$ octahedron rotational angle increases with decreasing temperature and reveals a rotation angle about 7.5(3)˚ and 8.1(2)˚ for 200 K and 90 K, respectively (see Table IV). There was no measured octahedral tilt (see Fig. 1(c)) in the bulk between 298 and 90 K.

Evidence for the rotation of $RuO_6$ octahedra on the (001) surface of $Sr_3Ru_2O_7$ was previously reported based on STM image[22]. However, STM imaging reflects the morphology of the surface density of states (DOS) instead of the lattices. Although this morphology is usually expected to follow the surface structure periodicity, detailed structural information cannot be extracted from STM images. Thus, LEED analysis was employed for a quantitative structure determination of $Sr_3Ru_2O_7$(001). The impinging low energy electrons strongly interact with the atoms at the top surface layers. This strong interaction gives rise to a multiple scattering process that reduces the free mean path of the probing electrons and enhances the surface sensitivity of the technique. Another consequence of this multiple scattering is that surface structure determination by LEED needs to follow an indirect methodology, in which the experimentally collected *I-V* curves are compared with theoretically calculated ones for a variety of structures. This comparison is made quantitatively by using the Pendry reliability factor ($R_P$)[27]. A low $R_P$ value suggests that the structural results are reliable.

A modified version of the symmetrized automated tensor LEED code (SATLEED)[28] was employed in our theoretical calculations[29]. Atomic phase shifts were calculated within the optimized muffin-tin (MT) potential approximation[30]. Debye temperatures for each element in $Sr_3Ru_2O_7$ were determined from the isotropic mean square displacements obtained from our XRD results. Fig. 2(a) shows the LEED diffraction pattern of $Sr_3Ru_2O_7$ (001) surface at 300 K after a fresh surface cleavage. For a bulk truncated (001) surface, the rotation of the bilayer octahedra will generate glide-lines in the LEED pattern which will produce extinguished diffracted spots. This is illustrated in Fig. 2(b), where spots labeled as ($\pm h$,0) and (0,$\pm h$) (h=1, 3, 5, …) are extinguished at all energies. The dashed lines represent the glide lines. As can be seen in Fig. 2(a), one of the glide lines for $Sr_3Ru_2O_7$(001) is absent, and spots such as (3,0) and (-3,0) are clearly visible. At subsequent cooling and warming cycle, spots (0,3) and (0,-3) were always absent, but spots (3,0) and (-3,0) were always sustained. Several cleavages from different sample batches reproduced this result. This implies a different symmetry at the surface produced by truncation of the bulk. This is very similar to the single-layered ruthenate surface[8,9], the symmetry consideration indicates that the absence of a glide line is due to a tilt of the top layer octahedra. In bulk $Sr_3Ru_2O_7$, the octahedra are rotated by an angle of ~ 6.7° at room temperature without any sign of tilt. *The surface presents a lower symmetry than bulk.* Fig. 2(c) illustrates the rotation and tilt angles of the $RuO_6$ octahedron at the surface as well as the surface unit cell. Its dimensions correspond to a (√2x√2)R45° unit cell compared with the bulk truncated one (1x1). This is because the surface unit cell takes into account the rotation of the octahedra. In the bulk unit cell (tetragonal – *I4/mmm*), the rotation is represented by the splitting of the O3 positions. Using such a surface unit cell, we can avoid labeling some diffracted beams with fractions, the latter are easily mistaken for the presence of superlattices due to surface reconstruction.

The structure determination of the (001) surface of $Sr_3Ru_2O_7$ was performed by employing a quantitative comparison between the experimentally and theoretically generated *I-V* curves. As mentioned above, the surface has a lower symmetry due to the tilt of top octahedral layer. The surface structure can be described by the plane group, *p2gg* (No. 8). As shown in Fig. 2(d), the surface atomic displacements were determined via the following steps: (1) after setting the top octahedra tilt angle (Θ) at 2°, the rotation angle (Φ) was optimized (grid search) in the theoretical model in order to minimize $R_P$ ($\Phi_{MIN}$); (2) using $\Phi=\Phi_{MIN}$, the tilt angle Θ was then optimized ($\Theta_{MIN}$) for minimum $R_P$; (3) with $\Theta_{MIN}$ and $\Phi_{MIN}$ fixed, the Sr1 ($\Delta Z_4$) and Sr2 ($\Delta Z_2$) vertical positions [along (001)] were optimized ($Sr1_{BEST}$,$Sr2_{BEST}$); (4) using $\Theta_{MIN}$, $\Phi_{MIN}$, $Sr1_{BEST}$ and $Sr2_{BEST}$, the motion of O2 ($\Delta Z_1$), Ru ($\Delta Z_3$) and O1 ($\Delta Z_5$) atoms was restricted along O2-Ru-O1 bonding direction and their relative positions were optimized in order to reduce $R_P$; (5) this procedure was repeated until the $R_P$ reached to a global minimum value. The associated errors in the structural parameters were calculated using the methodology discussed elsewhere[27,31]. Very good theory-experiment agreement was obtained for both 300 and 80 K data sets, as characterized by the final $R_P$ values of 0.27 and 0.22, respectively. Fig. 3 shows typical theoretical and experimental *I-V* curves for comparison at 80 K.

Our results for the top layer octahedra, presented in Table V, clearly indicate a lower symmetry at the surface than in the bulk. The top layer octahedra present a tilt of (4.5 ± 2.5°) at 300 K and (2.5 ± 1.7)° at 80 K, which was not observed in the bulk. Note that the surface octahedral rotation angle (~ 12°) does not statistically change within the temperature range between 80 and 300 K. This is in contrast to what occurs in the bulk, which has an increase in rotation angle as the temperature decreases (*i.e.* 6.7(6)° at 298 K to 8.1(2)° at 90 K).

In the prior neutron diffraction study of $Sr_3Ru_2O_7$, Shaked *et al.* reported a high degree of strain in the powder samples, which was attributed to the strong anisotropic thermal expansion. It was suggested that this strain was relieved during successive thermal cycles and characterized by an increase in the $c/\langle a \rangle$ ratio[19]. For confirmation, we also carried out the thermal-cycle experiment. A total of thirteen XRD data collections were acquired. The refinement for each data collection using the best fitting space group, tetragonal *I*4/*mmm*, resulted in $R_1$ = 0.028 to 0.040 and *S* (Goodness of Fit) = 1.22 to 1.34 indicating the high quality of our single crystals. For direct comparison to the previously reported thermal cycling effect on the lattice parameter ratio $c/\langle a \rangle$ (the unit cell parameter *c* divided by the average of the unit cell parameters *a* and *b*) in an orthorhombic space group, *Bbcb*, our tetragonal unit cell parameter *a* was multiplied by √2. Fig. 4(a) shows thermal-cycling dependence of $c/\langle a \rangle$ ratios over three cycles. While the ratio increases with decreasing temperature, the overlap of the error bars for all measurements at a fixed temperature demonstrates little change over the course of three thermal cycles. Plotted in Fig. 4(b) is our ratio $c/\langle a \rangle$ for T = 298 K and 200 K and the ratio $c/\langle a \rangle$ for 298 K from Ref. 19. Our ratios at room temperature are in very good agreement (*i.e*. within the error of our measurements) with the previously reported room temperature ratios. However, based on our results, we cannot conclude that this ratio is increasing per thermal cycle. In fact, the ratio from the end of the third thermal cycle at room temperature was the lowest ratio among all thirteen data collections, which includes the very first room temperature ratio prior to any thermal cycling.

If it were due to the strain effect, one would expect an impact on the surface structure. The creation of a clean surface by single crystal cleavage breaks the translational symmetry in the *c* direction. The atoms in the top atomic layers present lower coordination, thus generating a type of uni-axial pressure along *c* direction. If strain can be relieved through cooling cycles, the LEED should reflect the change of the surface structure. We performed LEED experiments through cooling and warming cycle on the $Sr_3Ru_2O_7$(001) surface. Since the tilt of $RuO_6$ octahedra destroys the glide line with visible (3,0) and (-3,0) spots, we can track the change in the tilt angle by evaluating beam intensities with temperature and thermal cycle. For a quantitative comparison, it is necessary to exclude the Debye-Waller effect in the intensities. An easy solution for this is to renormalize the intensity of (3,0) beam using another typical beam whose intensity is not sensitive to the tilt angle. In the case of $Sr_3Ru_2O_7$(001) surface, the (1,1) and (2,2) spots were used for the ratio calculation. The higher the ratio between (3,0) and (1,1) and the ratio between (3,0) and (2,2), the larger is the tilt angle of the top octahedra. Since the (3,0) and (-3,0) spots are only clearly visible within ~ 30 eV energy range, the following approach was used to calculate the referred ratios: (1) the intensity of beams (3,0) and (-3,0) was determined by integrating the *I-V* spectra in a 30 eV energy range, which defines one diffraction peak; (2) the intensities of (3,0) and (-3,0) spots were averaged and called $I_{(3,0)}$; (3) the same 30 eV energy range was used in the *I-V* peak integration for (1,1) and (2,2) beam groups, *i.e.*, (1,1), (-1,1), (1,-1), (-1,-1) and (2,2), (-2,2), (2,-2), (-2,-2), respectively; (4) an average was used to define the $I_{(1,1)}$ and $I_{(2,2)}$ intensities; (5) the previous four steps were repeated for different temperatures between 300 and 80 K through cooling and warming cycles; (6) the uncertainties in the intensities were obtained from the standard deviation of the average intensity from the individual beam intensities. The reversible temperature dependence of $I_{(3,0)}/I_{(1,1)}$ and $I_{(3,0)}/I_{(2,2)}$ ratios in the cooling and warming cycle are shown in Fig. 5(a) and (b), respectively. It was observed that both $I_{(3,0)}/I_{(1,1)}$ and $I_{(3,0)}/I_{(2,2)}$ slightly decrease from 300 to 80 K during cooling process and then slightly increases when warming up. This was consistent with our structural results previously discussed that show a smaller tilt angle at 80 K (see Table V). If strain were relieved in thermal cycles, it would be reflected in the intensity ratios. Our experimental data in Fig. 5(a) and (b) indicates that the ratios have little change between cooling-down and warming-up processes. Therefore, both our bulk and surface results show no evidence for the relieved strain.

In summary, we have performed an investigation of both the bulk and surface structures of $Sr_3Ru_2O_7$ single crystals. Our single-crystal XRD data is best modeled as a tetragonal structure (*I*4/*mmm*) with ~ 6.7° of $RuO_6$ octahedral rotation and a split occupancy of the equatorial oxygens, O3. On the other hand, the surface structure is quite distinct from the bulk: it is more distorted with the top $RuO_6$ octahedra not only rotated with higher angle than the bulk (~ 12°), but also tilted (~ 4.5°) at room temperature. This tilting structural phase was not observed for bulk between 90 and 298 K. The surface rotation angle remains constant, while the tilting angle slightly decreases as decreasing temperature between 80 and 300 K. Through thermal cycles between 300 and 80 K, no significant change in both surface and bulk structures.

Since lattice distortions are strongly coupled with the orbital and spin degrees of freedom in this material, the structural difference likely leads to different physical properties between surface and bulk. According to theoretical calculations[10] for single-layered $(Sr,Ca)_2RuO_4$, the rotation and tilt of $RuO_6$ octahedra are in favor of ferromagnetism and antiferromagnetism, respectively. The large rotation and tilting of $RuO_6$ octahedra at the surface of $Sr_3Ru_2O_7$ may imply strong competition between antiferromagnetic and ferromagnetic interactions. Experimental investigation of surface magnetism will be carried out.

**Acknowledgements**

We thank Frank Fronczek for technical assistance and useful discussion, and financial support of NSF DMR-0756281 (JYC) and NSF DMR-0451163 (BH, VBN and EWP).

**References:**

[*] Corresponding author: rjin@lsu.edu

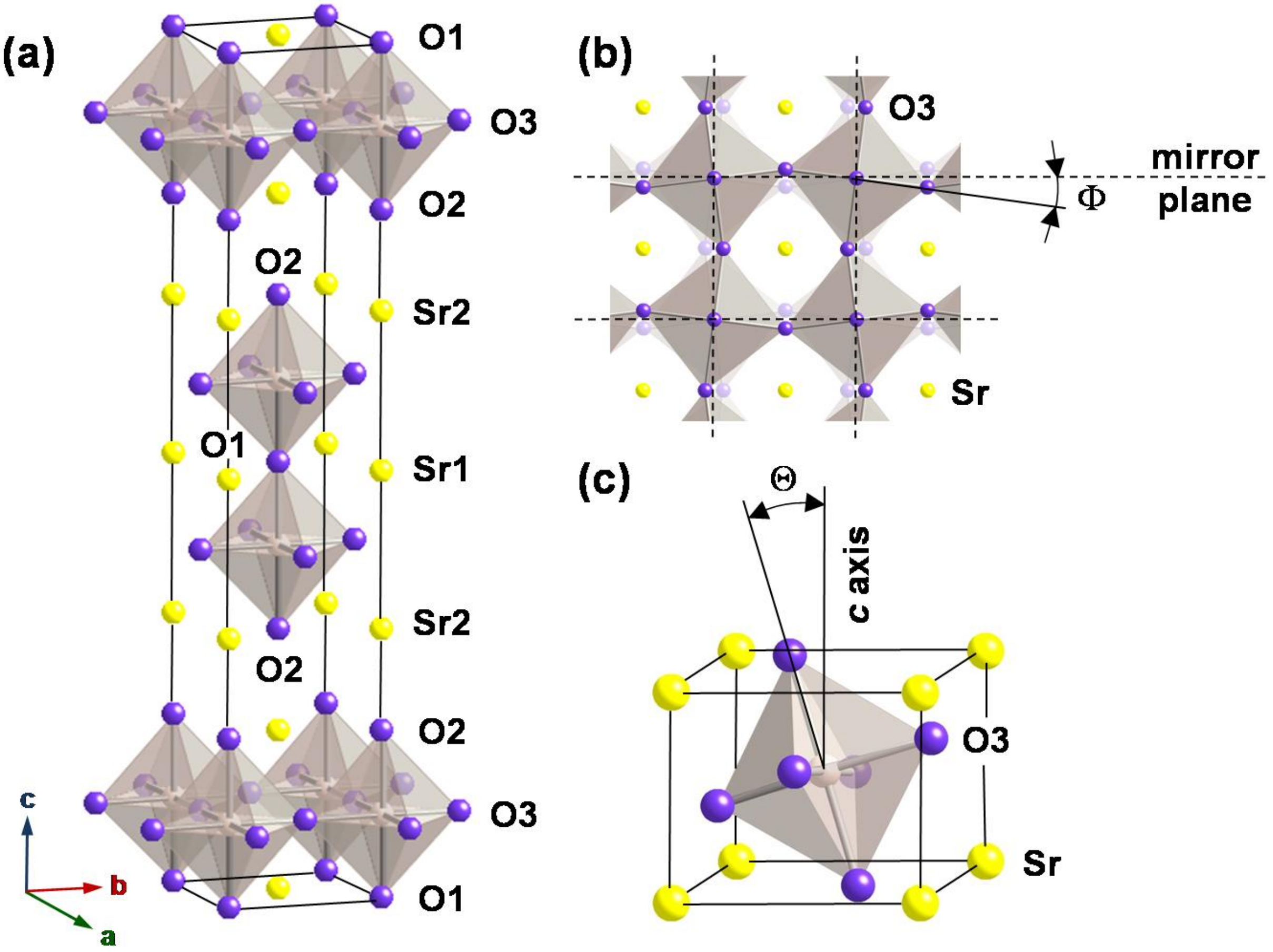

Fig. 1 (color online): (a) Unit cell representation of $Sr_3Ru_2O_7$ using space group *I*4/*mmm*. The Ru atoms are located in the center of each octahedron. (b) Top view of the $RuO_6$ octahedron showing the rotation angle (Φ) in the *ab*-plane (the dash lines present mirror planes). (c) View of the $RuO_6$ octahedron showing a tilt angle (Θ). For bulk, Θ = 0 (see text).

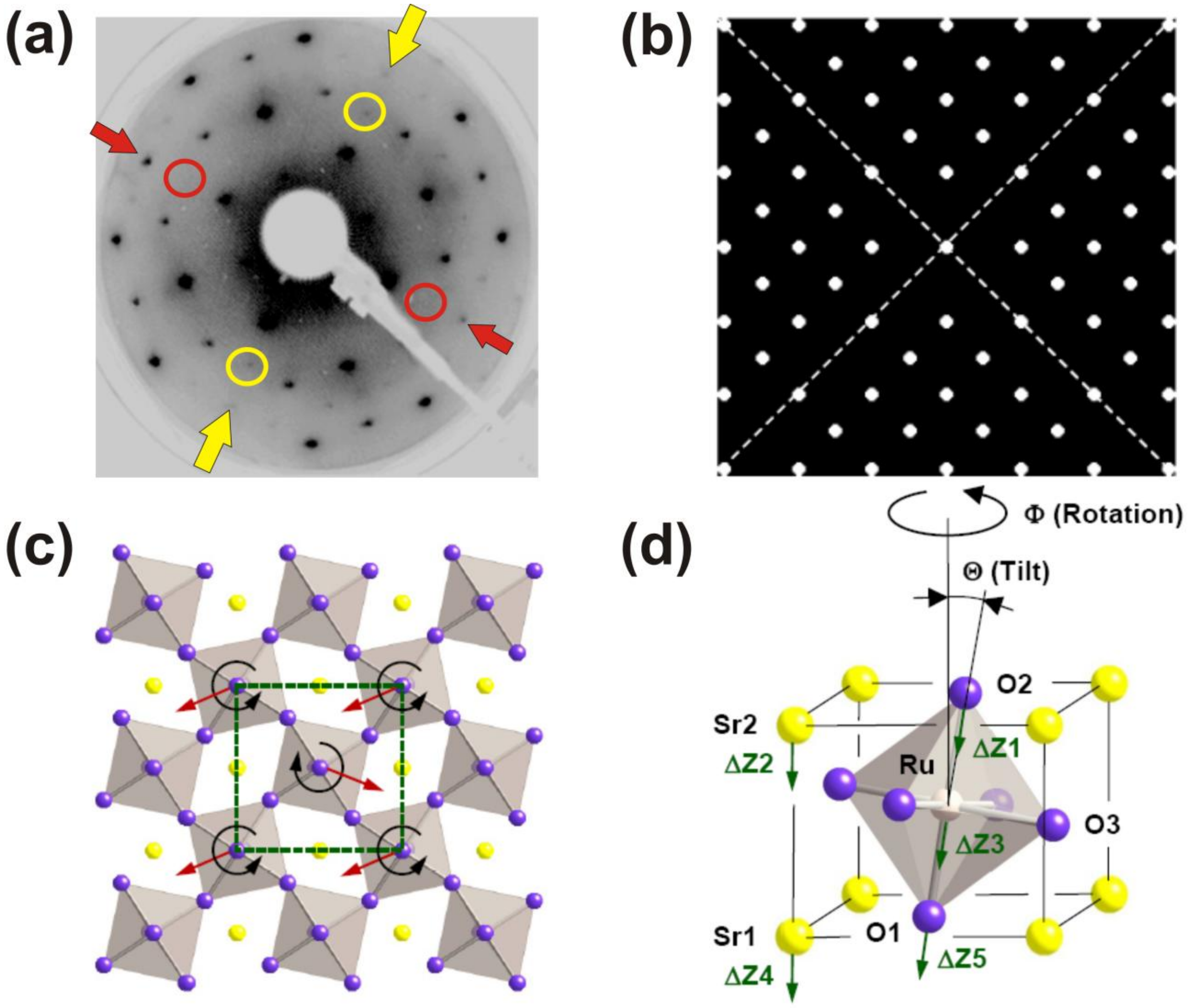

Fig. 2 (color online): (a) LEED diffraction pattern with energy of 225 eV at 300 K. The red arrows indicate the only existing glide line. The two red circles show the locations of the two extinguished spots [(0,3) and (0,-3)] along this line. Yellow arrows point the broken glide line, where (3,0) and (-3,0) diffracted spots are visible as indicated by the yellow circles. (b) Schematic diffraction pattern for a *p2gg* symmetry, with the two glide lines. (c) Top view of the (001) surface of $Sr_3Ru_2O_7$, in which the rotation (black arrows) and tilting (red arrows) of the octahedra can be visualized. The green dashed lines correspond to the surface unit cell. (d) Schematic illustration of the atomic displacements of $Sr_3Ru_2O_7$(001) surface (see Table 5).

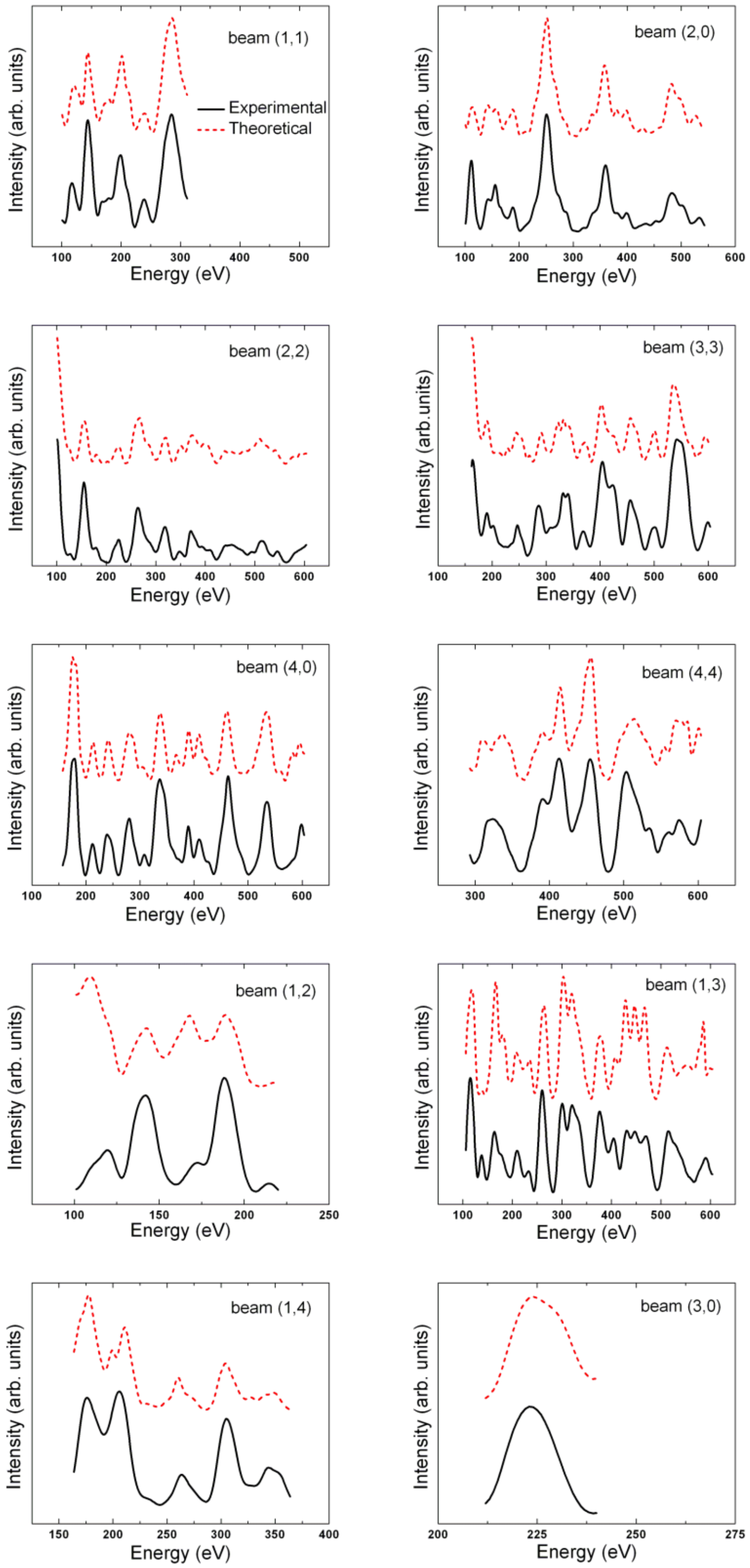

Fig. 3 (color online): Comparison between experimental and theoretically-generated *I(V)* curves for the final structure of $Sr_3Ru_2O_7$(001) surface at 80 K.

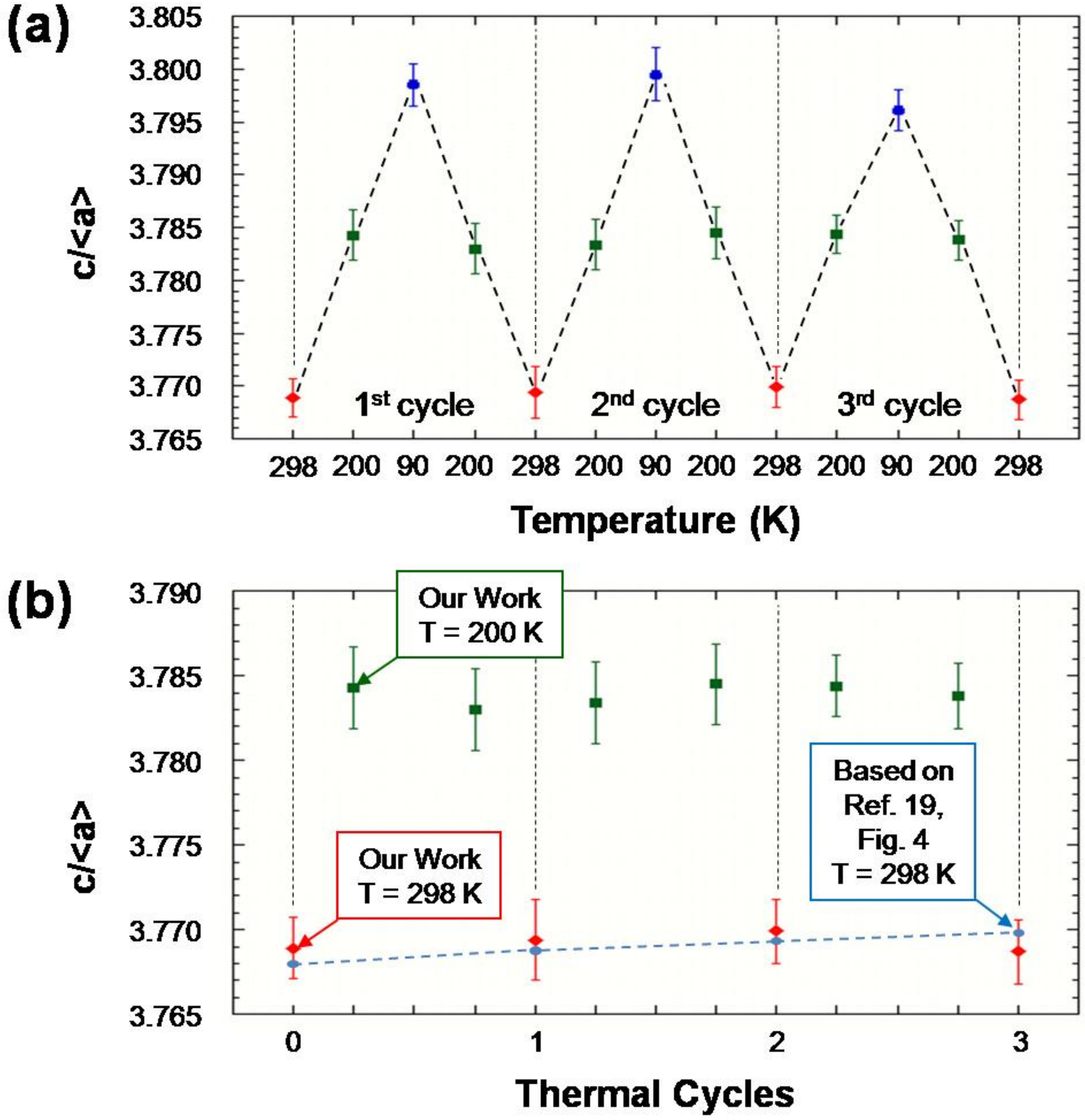

Fig. 4 (color online): (a) Temperature dependence of the cell parameter ratio c/<a>, from X-ray single crystal diffraction data, in three thermal cycles. For comparison to previously reported c/<a> thermal cycle data based on an orthorhombic unit cell, the tetragonal cell parameters *a* and *b* have been converted by multiplying by √2. The ratios are shown as filled diamonds for 298 K, filled squares for 200 K, and filled circles for 90 K. The diagonal dashed lines connecting the points are for guiding the eye. The vertical dashed lines are used to separate three cycles. (b) Cell parameter ratio c/<a> versus thermal cycles at 298 K (filled diamonds) and 200K (filled squares). For comparison, previously reported c/<a> ratios from powder neutron diffraction (see Ref. 19) are plotted (filled circles connected with a dashed line). In both (a) and (b), the vertical dashed lines are used to separate three cycles.

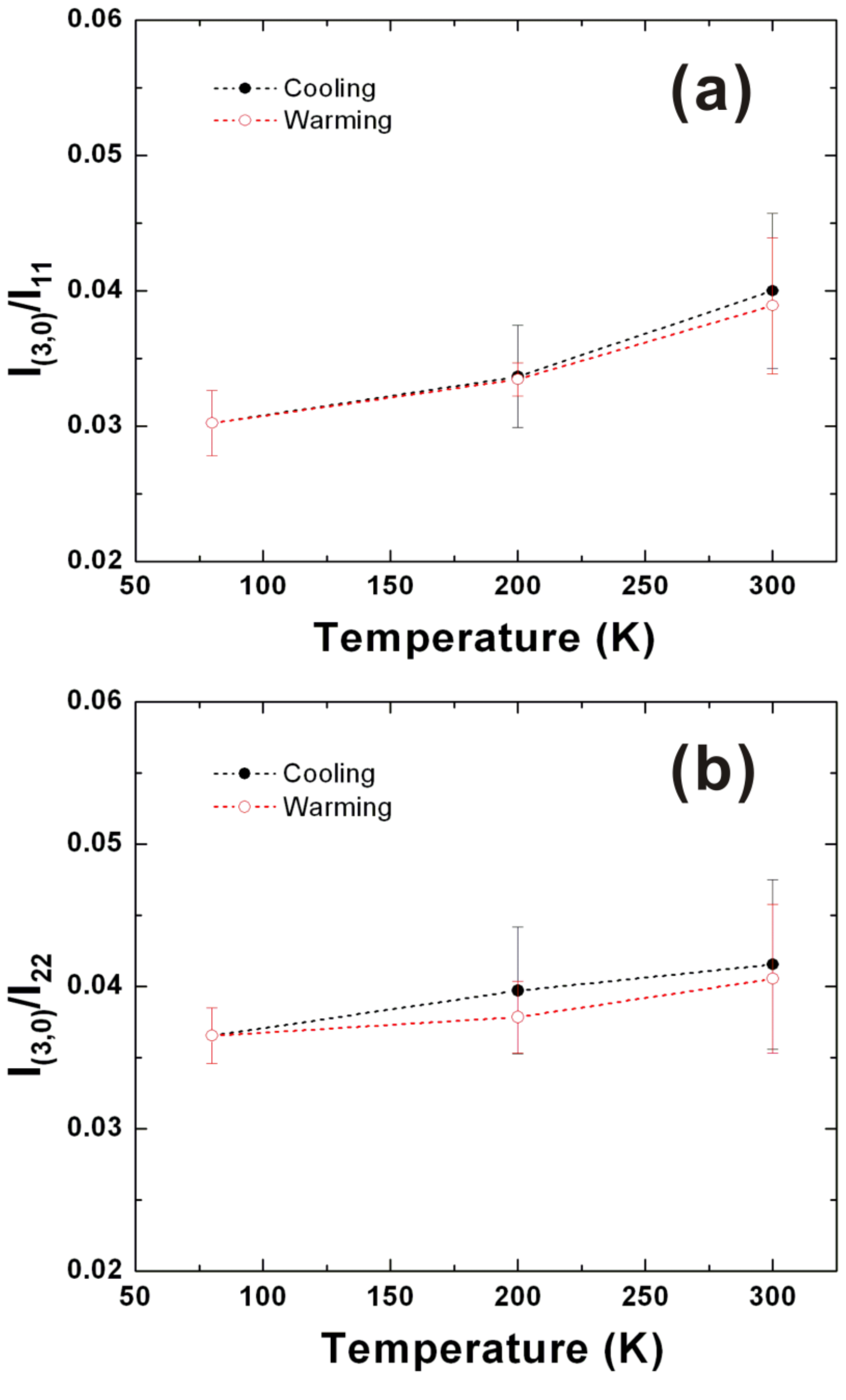

Fig. 5 (color online): (a) Temperature dependence of the intensity ratio between (3,0) and (1,1) diffracted spots via cooling (filled circles) and warming (empty circles). (b) Temperature dependence of the intensity ratio between (3,0) and (2,2) diffracted spots via cooling and warming. The dashed lines are guides for the eye.

## Table I: Crystallographic Parameters of $Sr_3Ru_2O_7$

| | | | |
|---|---|---|---|
| *Crystal data* | | | |
| Temperature (K) | 298(2) | 200(2) | 90(2) |
| Formula | $Sr_3Ru_2O_7$ | $Sr_3Ru_2O_7$ | $Sr_3Ru_2O_7$ |
| Crystal system | Tetragonal | Tetragonal | Tetragonal |
| Space group | *I*4/*mmm* (No. 139) | *I*4/*mmm* (No. 139) | *I*4/*mmm* (No. 139) |
| $a$ (Å) | 3.8897(10) | 3.8800(15) | 3.8716(10) |
| $c$ (Å) | 20.7320(60) | 20.7669(70) | 20.7980(80) |
| $V$ (Å$^3$) | 313.66(15) | 312.70(20) | 311.75(15) |
| Mosaicity (°) | 0.427(6) | 0.491(6) | 0.471(6) |
| Z | 2 | 2 | 2 |
| $2\theta$ range (°) | 7.86-54.72 | 7.84-54.88 | 7.84-54.68 |
| $\mu$ (mm$^{-1}$) | 30.032 | 30.136 | 30.215 |
| | | | |
| *Data collection* | | | |
| Measured reflections | 363 | 360 | 345 |
| Independent reflections | 142 | 142 | 140 |
| Reflections with $I>2\sigma(I)$ | 139 | 139 | 138 |
| $^aR_{int}$ | 0.0363 | 0.0605 | 0.0321 |
| $h$ | -5→5 | -4→5 | -5→5 |
| $k$ | -3→3 | -4→5 | -3→3 |
| $l$ | -26→23 | -24→26 | -24→26 |
| | | | |
| *Refinement* | | | |
| Reflections | 142 | 142 | 140 |
| Parameters | 21 | 21 | 21 |
| $^bR_1[F^2>2\sigma(F^2)]$ | 0.0300 | 0.0347 | 0.0318 |
| $^cwR_2(F^2)$ | 0.0878 | 0.0909 | 0.0802 |
| $^dS$ | 1.314 | 1.245 | 1.237 |
| $\Delta\rho_{max}$ (eÅ$^{-3}$) | 1.595 | 1.928 | 2.214 |
| $\Delta\rho_{min}$ (eÅ$^{-3}$) | -1.077 | -1.739 | -1.695 |

$^aR_{int} = [\ \Sigma\ |\ F_o^2 - F_c^2\ (\text{mean})\ |\ /\ (n\text{-}p)\ ]^{1/2}$

$^bR_1 = \Sigma\ |\ |\ F_o\ | - |\ F_c\ |\ |\ /\ \Sigma\ |\ F_o\ |$

$^cwR_2 = [\ \Sigma\ [\ w(\ F_o^2 - F_c^2\ )^2\ ]\ /\ \Sigma\ [\ w(\ F_o^2\ )^2\ ]\ ]^{1/2}$,
$w = 1\ /\ [\ \sigma^2(F_o^2) + (0.0495P)^2 + 0.6876P\ ]$ for 298 K,
$w = 1\ /\ [\ \sigma^2(F_o^2) + (0.0573P)^2 + 0.0000P\ ]$ for 200 K,
$w = 1\ /\ [\ \sigma^2(F_o^2) + (0.0525P)^2 + 0.0172P\ ]$ for 90 K

$^dS = [\ \Sigma\ [\ w(\ F_o^2 - F_c^2\ )^2\ ]\ /\ \Sigma\ (n - p)\ ]^{1/2}$

**Table II: Atomic Positions and Equivalent Isotropic Displacement Parameters**

| Atom | Wyckoff position | *x* | *y* | *z* | Occ.[a] | $U_{eq}$ (Å$^2$)[b] |
|---|---|---|---|---|---|---|
| **T = 298 K** | | | | | | |
| Sr1 | 2*b* | 1/2 | 1/2 | 0 | 1 | 0.0071(6) |
| Sr2 | 4*e* | 1/2 | 1/2 | 0.18626(9) | 1 | 0.0070(6) |
| Ru1 | 4*e* | 0 | 0 | 0.09741(5) | 1 | 0.0034(6) |
| O1 | 2*a* | 0 | 0 | 0 | 1 | 0.013(3) |
| O2 | 4*e* | 0 | 0 | 0.1958(5) | 1 | 0.013(2) |
| O3 | 16*n* | 1/2 | 0.059(5) | 0.0964(3) | 0.5 | 0.013(5) |
| **T = 200 K** | | | | | | |
| Sr1 | 2*b* | 1/2 | 1/2 | 0 | 1 | 0.0050(6) |
| Sr2 | 4*e* | 1/2 | 1/2 | 0.18634(8) | 1 | 0.0049(6) |
| Ru1 | 4*e* | 0 | 0 | 0.09740(5) | 1 | 0.0023(6) |
| O1 | 2*a* | 0 | 0 | 0 | 1 | 0.008(3) |
| O2 | 4*e* | 0 | 0 | 0.1963(5) | 1 | 0.0092(18) |
| O3 | 16*n* | 1/2 | 0.066(3) | 0.0967(2) | 0.5 | 0.010(3) |
| **T = 90 K** | | | | | | |
| Sr1 | 2*b* | 1/2 | 1/2 | 0 | 1 | 0.0030(5) |
| Sr2 | 4*e* | 1/2 | 1/2 | 0.18659(8) | 1 | 0.0032(5) |
| Ru1 | 4*e* | 0 | 0 | 0.09743(5) | 1 | 0.0017(5) |
| O1 | 2*a* | 0 | 0 | 0 | 1 | 0.005(2) |
| O2 | 4*e* | 0 | 0 | 0.1958(5) | 1 | 0.0062(16) |
| O3 | 16*n* | 1/2 | 0.0707(18) | 0.0969(2) | 0.5 | 0.008(3) |

[a] Occupancy of atoms

[b] $U_{eq}$ is defined as one-third of the trace of the orthogonalized $U^{ij}$ tensor.

**Table III: Anisotropic Atomic Displacement Parameters (Å$^2$)**

| Atom | $U^{11}$ | $U^{22}$ | $U^{33}$ | $U^{12}$ | $U^{13}$ | $U^{23}$ |
|---|---|---|---|---|---|---|
| **T = 298 K** | | | | | | |
| Sr1 | 0.0077(7) | 0.0077(7) | 0.0059(11) | 0.000 | 0.000 | 0.000 |
| Sr2 | 0.0080(6) | 0.0080(6) | 0.0050(10) | 0.000 | 0.000 | 0.000 |
| Ru1 | 0.0038(6) | 0.0038(6) | 0.0025(8) | 0.000 | 0.000 | 0.000 |
| O1 | 0.018(4) | 0.018(4) | 0.003(7) | 0.000 | 0.000 | 0.000 |
| O2 | 0.016(3) | 0.016(3) | 0.005(4) | 0.000 | 0.000 | 0.000 |
| O3 | 0.006(4) | 0.021(17) | 0.013(4) | 0.000(3) | 0.000 | 0.000 |
| **T = 200 K** | | | | | | |
| Sr1 | 0.0058(7) | 0.0058(7) | 0.0035(11) | 0.000 | 0.000 | 0.000 |
| Sr2 | 0.0057(6) | 0.0057(6) | 0.0033(10) | 0.000 | 0.000 | 0.000 |
| Ru1 | 0.0027(6) | 0.0027(6) | 0.0015(8) | 0.000 | 0.000 | 0.000 |
| O1 | 0.008(3) | 0.008(3) | 0.009(7) | 0.000 | 0.000 | 0.000 |
| O2 | 0.011(2) | 0.011(2) | 0.005(4) | 0.000 | 0.000 | 0.000 |
| O3 | 0.006(3) | 0.013(11) | 0.011(4) | -0.001(2) | 0.000 | 0.000 |
| **T = 90 K** | | | | | | |
| Sr1 | 0.0038(6) | 0.0038(6) | 0.0016(10) | 0.000 | 0.000 | 0.000 |
| Sr2 | 0.0038(5) | 0.0038(5) | 0.0019(9) | 0.000 | 0.000 | 0.000 |
| Ru1 | 0.0025(6) | 0.0025(6) | 0.0003(8) | 0.000 | 0.000 | 0.000 |
| O1 | 0.006(3) | 0.006(3) | 0.002(6) | 0.000 | 0.000 | 0.000 |
| O2 | 0.009(2) | 0.009(2) | 0.000(4) | 0.000 | 0.000 | 0.000 |
| O3 | 0.007(3) | 0.010(8) | 0.006(4) | -0.001(2) | 0.000 | 0.000 |

**Table IV: Selected Bond Distances (Å) and Angles (°)**

| | 298K | 200K | 90K |
|---|---|---|---|
| *Distances* | | | |
| Sr1-O1 | 2.75065(14) | 2.7436(2) | 2.73792(14) |
| Sr1-O3 (×4) | 2.635(15) | 2.621(7) | 2.613(6) |
| Sr2-O2 | 2.445(11) | 2.438(11) | 2.7446(7) |
| Sr2-O3 (×4) | 2.534(15) | 2.512(8) | 2.891(6) |
| Ru1-O1 | 2.0195(11) | 2.0227(11) | 2.0263(10) |
| Ru1-O2 | 2.040(10) | 2.053(11) | 2.046(9) |
| Ru1-O3 (×4) | 1.958(3) | 1.9566(13) | 1.9553(10) |
| *Angles* | | | |
| O1-Ru1-O3 (×4) | 89.39(19) | 89.56(14) | 89.70(15) |
| O2-Ru1-O3 (×4) | 90.61(19) | 90.44(14) | 90.30(15) |
| *Rotation* | | | |
| [a]$RuO_6$ octahedra | 6.7(6) | 7.5(3) | 8.1(2) |

[a] This value represents the rotational angle ($\Phi$) for the $RuO_6$ octahedra (see Fig. 1b).

**Table V:** Structural parameters of $Sr_3Ru_2O_7$(001) surface at 300 and 80 K, including the displacements of the atoms on the surface layer with respect to a bulk truncated structure as determined by our X-ray measurements. Bulk values for the octahedra rotation and tilt angles as well as for the Ru-O distances are also presented for comparison.

| Parameter | 300 K | 80 K |
|---|---|---|
| ΔZ1 (O(2)) | (0.040±0.060) Å | (0.060 ± 0.040) Å |
| ΔZ2 (Sr top) | (0.020±0.020) Å | (0.050 ± 0.015) Å |
| ΔZ3 (Ru) | (-0.010±0.020) Å | (0.025 ± 0.020) Å |
| ΔZ5 (O(1)) | (-0.020±0.080) Å | (0.015 ± 0.040) Å |
| ΔZ4 (Sr middle) | (0.020±0.030) Å | (0.045 ± 0.015) Å |
| Ru-O(2) | (1.990 ± 0.040) Å | (2.011 ± 0.030) Å |
| | Bulk: 2.0400 Å | Bulk: 2.0460 Å |
| Ru-O(3) | (1.988 ± 0.035) Å | (1.979 ± 0.025) Å |
| | Bulk: 1.9580 Å | Bulk: 1.9553 Å |
| Ru-O(1) | (2.009 ± 0.050) Å | (2.016 ± 0.030) Å |
| | Bulk: 2.0195 Å | Bulk: 2.0263 Å |
| $RuO_6$ rotation | (12 ± 5)° | (12 ± 3)° |
| | Bulk: 6.7° | Bulk: 8.1° |
| $RuO_6$ tilt | (4.5 ± 2.5)° | (2.5 ± 1.7)° |
| | Bulk: no tilt | Bulk: no tilt |
| Rp | (0.27 ± 0.03) | (0.22 ± 0.02) |